\RequirePackage{fix-cm}
\documentclass[smallextended]{svjour3}       
\smartqed  
\usepackage{graphicx}
\usepackage{amsmath}
\usepackage{booktabs}
\usepackage{multirow}
\usepackage{amssymb}
\usepackage{textcomp}

\begin{document}

\title{Simulation of a Laue lens with bent Ge(111) crystals}


\author{Vineeth Valsan \and Enrico Virgilli \and Filippo Frontera \and Vincenzo Liccardo \and Ezio Caroli \and John B Stephen}

\institute{V. Valsan \and E. Virgilli \and F. Frontera \and V. Liccardo \at
              Department of Physics, University of Ferrara, Via Saragat 1/c, 44122 Ferrara, Italy \\
              \email{valsan@fe.infn.it}\\       
            \emph{Present address:} of V. Valsan \at
            Indian Institute of Astrophysics, Bangalore 560034, India \\
            \email{vineeth@iiap.res.in}\\ 
            \and	  
           V. Valsan \and V. Liccardo \at
           Universit\'e de Nice Sophia-Antipolis, Parc Valrose, 06108 Nice Cedex 2, France \\     
           \and
           E. Caroli \and J. B. Stephen \at
           IASF-Bologna, INAF, Via Gobetti 101, 40129 Bologna, Italy\\
}

\date{Received: date / Accepted: date}

\maketitle

\begin{abstract}
In the context of Laue project for focusing hard X-/ soft gamma-rays, an entire Laue lens, using bent Ge(111) crystal tiles, with 40 meters curvature
radius, is simulated with a focal length of 20 meters.
The focusing energy band is between 80 keV and 600 keV.
The distortion of the output image of the lens on the focal plane due to the effect of crystal tile misalign.ment as well as the radial distortion
arising from the curvature of the crystal is discussed in detail.
Expected detection efficiency and instrument background is also estimated. Finally the sensitivity of the Laue lens is
calculated. A quantitative analysis of the results of these simulation is also presented.

\keywords{Focusing telescopes \and X-ray diffraction \and Laue lens \and Experimental astronomy \and High energy instrumentation}
\end{abstract}

\section{Introduction}
\label{intro}
Motivated by the astrophysical importance \cite{Frontera13} of extending the focusing band  up to 600 keV, 
a project named "LAUE" (supported by the Italian Space Agency)
was started with the goal of building Laue lenses\cite{Frontera11} for broad energy band (70-600 keV).
Previously, within the project named HAXTEL 
(Hard X-ray Telescope), two prototypes of Laue lenses made of flat mosaic crystals with short focal length (6m)
have been developed and successfully tested \cite{Frontera12,Frontera12b,Virgilli11,Virgilli11b,Valsan11}.
In the framework of LAUE project, a  prototype of  lens petal made of curved crystals is being developed in the LARIX 
facility of the Physics and Earth Sciences Department of the University of Ferrara.

Initially, a petal structure of the lens was simulated and built \cite{Liccardo12,Valsan12}.
The simulations made for the petal have been extended to model a complete Laue lens, made of bent Ge(111) crystals,
with an energy passband from 90 -- 600 keV. Preliminary results have been reported in \cite{Liccardo13,Valsan13,Virgilli13,Valsan13b}.
Bent crystals, when compared to their flat mosaic counterparts, have higher efficiency and very low mosaicity,
i.e., curved crystals have high diffraction efficiency and a better capability of concentrating
the signal collected over a large area into a small focal spot. 
In this paper, the details of this simulation with the results will be discussed.

\section{Simulating the entire Laue lens}

The simulation uses bent Ge(111) crystal tiles with radius of curvature equal to 40 meters 
(i.e. twice the lens focal length) and the dimension of each tile is 30$\times$10$\times$2 mm$^3$. 
The quasi-mosaic crystal has a mosaicity of 4.12 arcsec.
The bending technology  adopted for Ge is the surface grooving \cite{Guidi11}.
From the dynamical theory of diffraction of bent crystals, it results that the bending of a perfect crystal 
creates a curvature of the lattice planes (111) of Ge, 
with a ratio between the internal curvature and external curvature radii of 2.39 \cite{Bellucci13}.
This gives rise to a quasi-mosaic configuration of the bent crystal. This effect is not valid for all lattice planes. 
For example, if the diffracting planes in transmission configuration are (220), 
the quasi-mosaicity is not created.

The advantages offered by bent crystals are that their reflectivity  exceeds the theoretical limit of 50\%, 
valid for flat perfect and mosaic crystals. 
The focusing capability is also better when the crystal is bent \cite{Guidi11b}. 
Experimental tests confirm the expectations \cite{Liccardo12}.

The simulation of the entire Laue lens provides the space distribution of the focused photons in the focal plane (Point Spread Function, PSF) for on--axis photons.  
It takes into account \\
-- the ideal case (without any distortions or misalignments); \\ 
-- different values of misalignment (in arcsec) in the positioning of the crystal tile on the lens frame and radial distortions (in meters) from the required curvature radius of 40 meters.

This investigation will help us in determining the acceptable level of distortion to get a reasonable PSF.

\subsection{Effect of tile misalignment}

Each crystal tile is positioned on the lens petal frame with an orientation
such that the image of the beam reflected by the crystal is formed on the lens focus, where there is
the position sensitive focal plane detector.

\begin{figure}[!ht]
\centering
\begin{minipage}{.65\textwidth}
    \centering
    \includegraphics[scale=0.25]{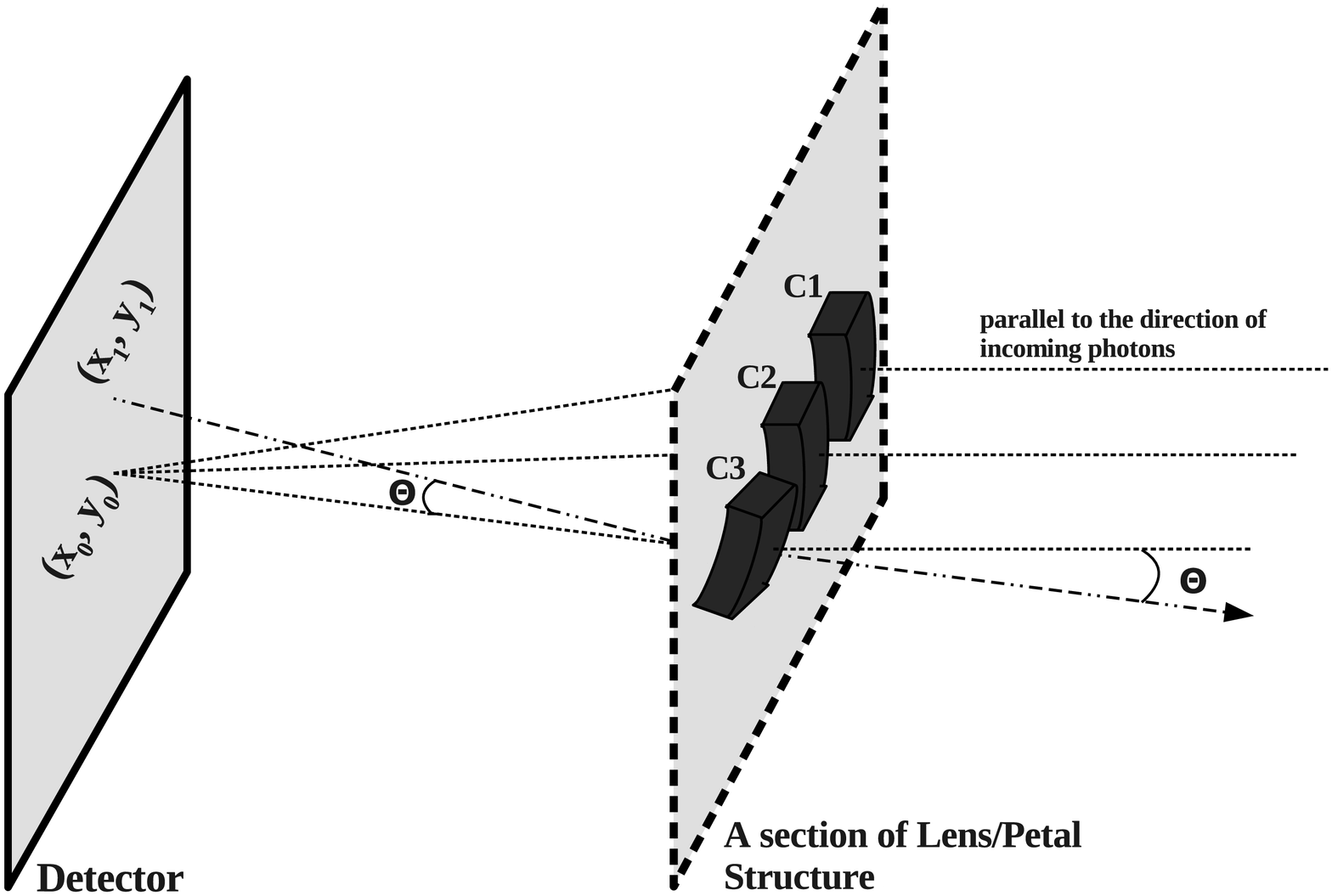}
\end{minipage}%
\begin{minipage}{.35\textwidth}
    \centering
    \includegraphics[scale=0.25]{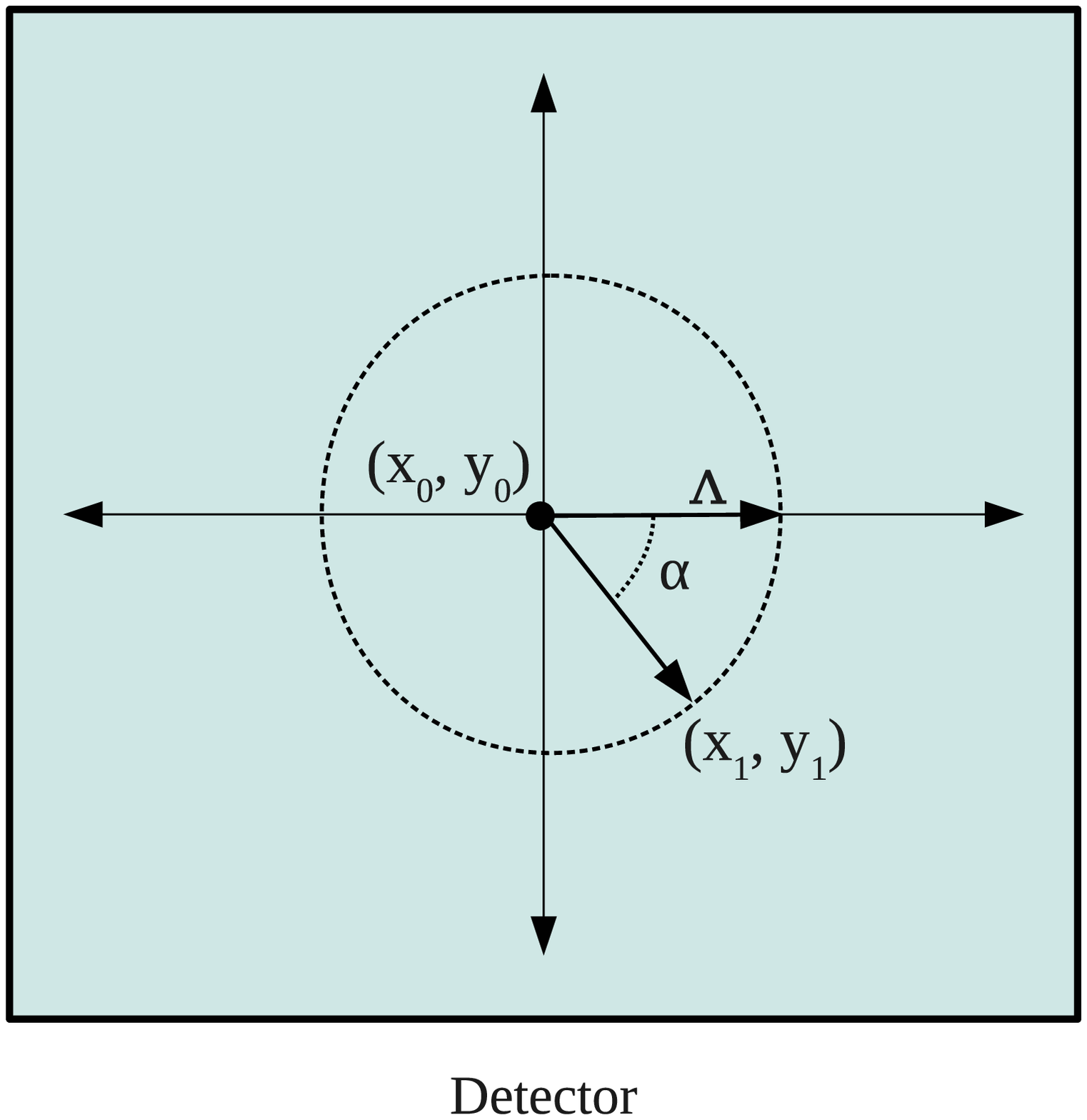} 
\end{minipage}
\caption{Misalignment in the positioning of the crystals on the lens petal frame 
deviates the PSF. $Left$ figure illustrates this effect. $(x_0, y_0)$ is the position
on the detector where the center of the image formed by the reflection of X-rays through perfectly
positioned crystals(C1 \& C2) falls. When there is a misalignment of $\Theta$ arcsec in the positioning (crystal C3), 
the center of the image is formed at $(x_1, y_1)$. The figure on the $right$ shows the focal plane 
(where there is the detector) with these positions.}
\label{fig:misalignment1}
\end{figure}

When a crystal is having a misalignment from its proper position, the image of that crystal
on the detector will have a deviation. In Fig.~\ref{fig:misalignment1}, let $(x_0, y_0)$ be the 
center of the image formed by the reflection of rays from a crystal, which is 
perfectly positioned on the lens petal frame without any misalignment.
$(x_1, y_1)$ is the position of the center of the image formed by a crystal having
a misalignment of $\Theta$ arcsec with respect to its proper positioning.
Crystals $C1$ and $C2$ are perfectly positioned, while crystal $C3$ have a misalignment of
$\Theta$ arcsec. The center of its image is formed on the detector at $(x_1, y_1)$.
The deviation of the center of the image from $(x_0, y_0)$ to $(x_1, y_1)$ is shown on the right side 
of Fig.~\ref{fig:misalignment1}. For a given misalignment of $\Theta$ arcsec, the linear deviation $\Lambda$ is given by:

\begin{equation}
 \Lambda = f~\Theta
\end{equation}

where $f$ ( = 20 meters) is the focal length of the lens.

Given that there is a uniform probability of having the point $(x_1, y_1)$ over the locus of the circle with center at
$(x_0, y_0)$ and a radius of $\Lambda$, the angle $\alpha$ (see Fig.~\ref{fig:misalignment1}, \textit{right})
is expected to have a uniform distribution.

The expected misalignment, $\Theta$, is of the order of few arcsec.
The effects of this misalignment have been incorporated in the code, 
with a uniform distribution.

\subsection{Effect of radial distortion}

\begin{figure}[h]
  \centering
  \includegraphics[scale=0.25]{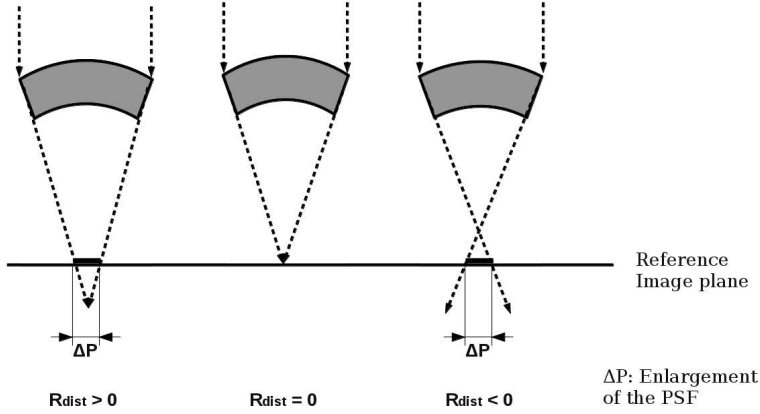}
  \caption{The effect of radial distortion ($R_{dist}$ or $R_d$) on the PSF of the lens.}
  \label{fig:radialDisto}
\end{figure}

Figure \ref{fig:radialDisto} illustrates the effect of radial distortion ($R_{dist}$) 
on the PSF. The width of the PSF is equal for a symmetrical deviation of the
curvature radius from its expected value. That is, if $R_p$ is the expected value of the
curvature radius, and $\Delta R$ is the distortion from the expected value, the width of the 
PSF is equal for a curvature radius of $R_p + \Delta R$ or $R_p - \Delta R$.

\begin{figure}[h!]
  \centering
  \includegraphics[scale=0.4,keepaspectratio=true]{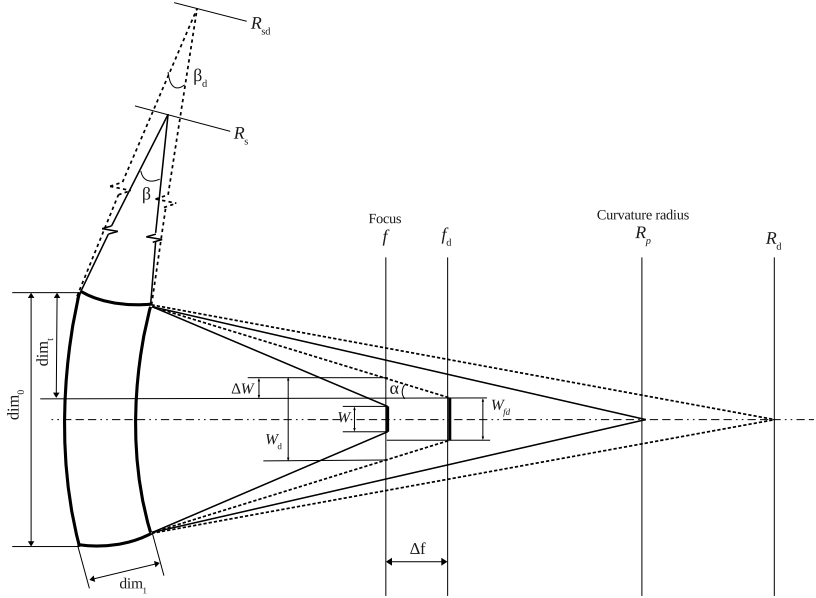}
  \caption{Diagram illustrating the effect of radial distortion of bent Ge(111) crystal
  on the width of the PSF.}
  \label{fig:Radial_Disto_Ge}
\end{figure}

In Fig.~\ref{fig:Radial_Disto_Ge}, $R_p$ is the expected primary curvature radius,
$f (= 2R_p)$ is the corresponding focal length.
$R_s = 2.39 R_p$, is the secondary curvature radius corresponding to $R_p$.
$R_{dist} = R_p + \Delta R$, $\Delta R$ being the distortion in curvature radius.
$f_d$ is the focal length corresponding to $R_{dist}$.
$R_{sd} = 2.39 R_{dist}$, is the secondary curvature radius corresponding to $R_{dist}$.\\
At the focal distance, $f$, the width $W$ of the PSF is given by the angle $\beta$ of the secondary curvature (see Fig.~\ref{fig:Radial_Disto_Ge}):
\begin{align*}
      &W = f \dfrac{dim_1}{R_s} = f\beta
\end{align*}

Similarly at $f_d$, the width $W_{fd}$ is given by $W_{fd} = \beta_d f_d$,
where $\beta_d = \tfrac{dim_1}{R_{sd}}$. 
$W_d$ is the width of the image formed on the detector placed on the focal plane at a distance $f$,
and can be derived as follows.

From Figure \ref{fig:Radial_Disto_Ge},

\begin{align*}
    &dim_t = \dfrac{dim_0 - W_{fd}}{2}\\
    &\alpha = \dfrac{dim_t}{f_d}\\
    &\Delta f = |f_d - f|\\
    &\Delta W = \Delta f \alpha \\
    &W_d = W_{fd} + 2\Delta W
\end{align*}

For the statistical approach in the code, the same methodology used for the misalignment effect 
has been adopted also for the radial distortion. The range of radial distortion is assumed to be uniformly
distributed over all the crystals.
For example, for a maximum radial distortion of 6 meters, the code incorporates this effect with
a uniform distribution of the deviation of the radius in the range [0, 6) meters 
from the expected curvature radius of 40 meters.

\begin{table}[h]
  \caption{Ring-by-ring characteristics of the lens made by Ge(111) crystal tiles. }
  \vspace{0.3 cm}
  \begin{tabular}{ c  c  c  c  c  c }
    \toprule
    \multirow{2}{*}{Ring} & Radius & No. of crystal & Min. Energy & Max. Energy & Energy range \\ 
			  & (cm)   &    tiles	   &(keV)	 & (keV)       & (keV)        \\ \midrule
      1  & 12.673  & 79   &  535.609 & 679.409 & 143.799 \\ 
      2  & 15.673  & 98   &  442.047 & 535.608 & 93.560  \\ 
      3  & 18.673  & 117  &  376.312 & 442.047 & 65.734  \\ 
      4  & 21.673  & 136  &  327.596 & 376.311 & 48.714  \\ 
      5  & 24.673  & 155  &  290.048 & 327.596 & 37.547  \\ 
      6  & 27.673  & 173  &  260.222 & 290.047 & 29.825  \\ 
      7  & 30.673  & 192  &  235.958 & 260.221 & 24.263  \\ 
      8  & 33.673  & 211  &  215.833 & 235.957 & 20.124  \\ 
      9  & 36.673  & 230  &  198.871 & 215.833 & 16.961  \\ 
      10 & 39.673  & 249  &  184.381 & 198.871 & 14.489  \\ 
      11 & 42.673  & 268  &  171.859 & 184.381 & 12.521  \\ 
      12 & 45.673  & 286  &  160.930 & 171.859 & 10.928  \\ 
      13 & 48.673  & 305  &  151.308 & 160.929 & 9.621  \\ 
      14 & 51.673  & 324  &  142.771 & 151.307 & 8.535  \\ 
      15 & 54.673  & 343  &  135.146 & 142.770 & 7.624  \\ 
      16 & 57.673  & 362  &  128.295 & 135.146 & 6.850  \\ 
      17 & 60.673  & 381  &  122.104 & 128.294 & 6.189  \\ 
      18 & 63.673  & 400  &  116.484 & 122.104 & 5.619  \\ 
      19 & 66.673  & 418  &  111.358 & 116.483 & 5.125  \\ 
      20 & 69.673  & 437  &  106.664 & 111.357 & 4.693  \\ 
      21 & 72.673  & 456  &  102.350 & 106.663 & 4.313  \\ 
      22 & 75.673  & 475  &  98.371  & 102.349 & 3.977  \\ 
      23 & 78.673  & 494  &  94.690  & 98.371  & 3.680  \\ 
      24 & 81.673  & 513  &  91.275  & 94.690  & 3.414  \\ 
      25 & 84.673  & 532  &  88.097  & 91.274  & 3.176  \\ 
      26 & 87.673  & 550  &  85.134  & 88.097  & 2.963  \\ 
      27 & 90.673  & 569  &  82.363  & 85.133  & 2.770  \\
      28 & 93.673  & 588  &  79.767  & 82.362  & 2.595  \\ 
      \bottomrule
  \end{tabular}
  \label{tab:Ge_lens_rings}
\end{table}

Ring-by-ring details of the lens are given 
in Table~\ref{tab:Ge_lens_rings}. Different parameters and their corresponding values are
given in Table~\ref{tab:Ge_lens_parameter}.

\begin{table}[!h]
  \begin{center}  
  \caption{Parameters of the lens made by Ge(111) crystal tiles. }
  \vspace{0.3 cm}
    \begin{tabular}{ l  l }
    \toprule
    Parameter 		& Value 	\\ \midrule
    Focal length	& 20 meters 	\\
    Energy range	& 80 - 600 keV	\\
    Subtended angle	& 18 degree 	\\
    No. of Rings	& 28		\\
    Minimum radius	& 12.67 cm	\\
    Maximum radius	& 93.67 cm 	\\
    No. of crystal tiles & 9341	\\
    Crystal material 	& Ge(111)	\\
    Crystal dimension 	& 30 mm $\times$ 10 mm $\times$ 2 mm \\
    Crystal mass (total) & 2.07 g $\times$ 9341 = 19.335 kg\\
    \bottomrule
    \end{tabular}
    \label{tab:Ge_lens_parameter}
  \end{center}	
\end{table}

\begin{figure}[!ht]
\centering
\begin{minipage}{.5\textwidth}
    \centering
    \includegraphics[scale=0.45]{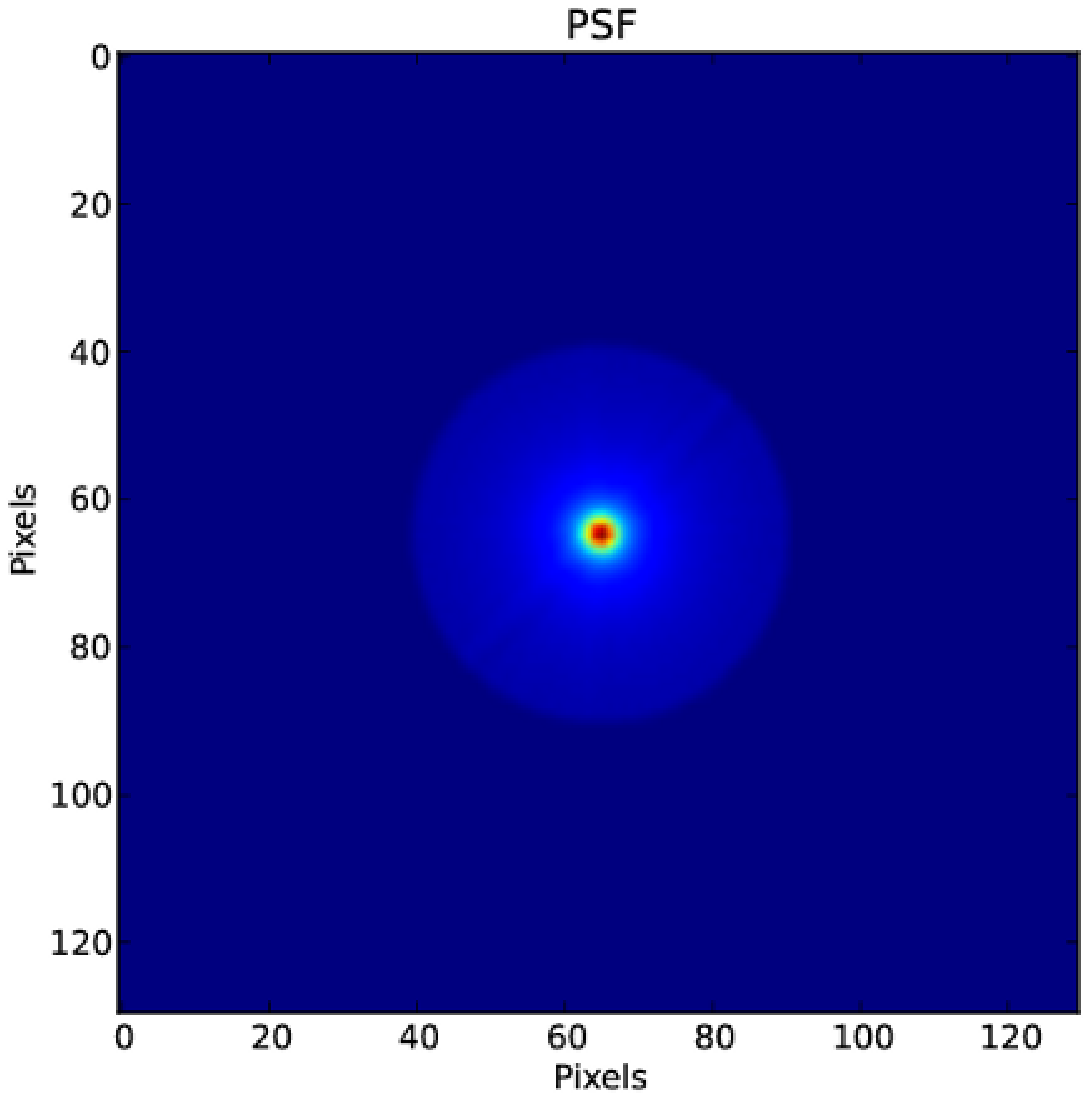}
    \renewcommand{\figure}{Fig.}
    
\end{minipage}%
\begin{minipage}{.5\textwidth}
    \centering
    \includegraphics[scale=0.65]{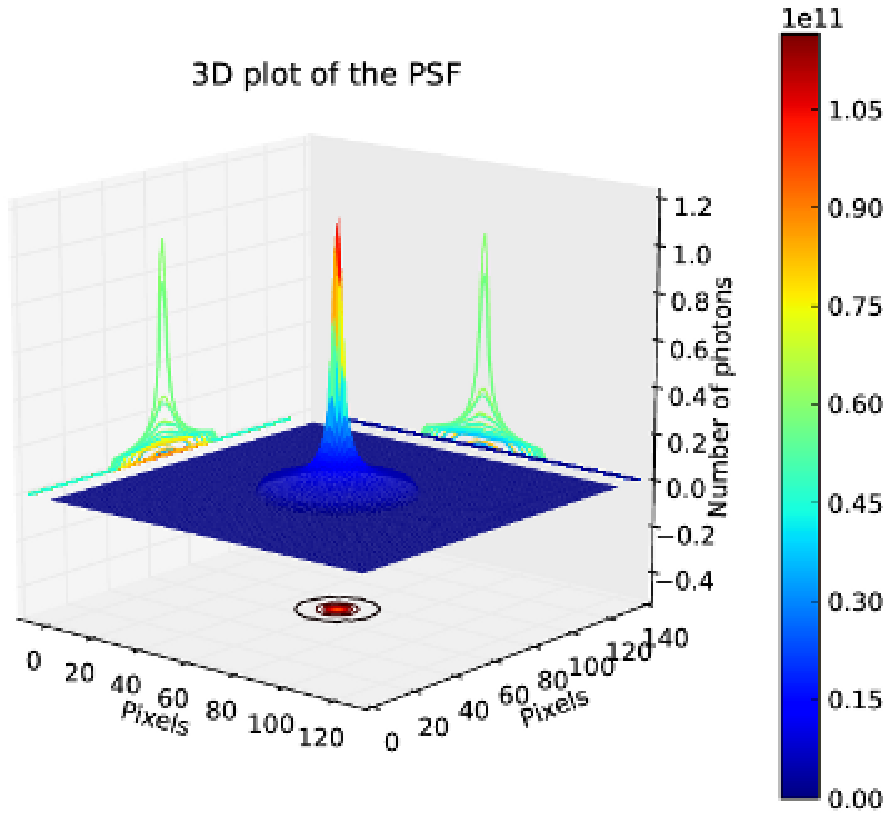} 
    \renewcommand{\figure}{Fig.}
    
\end{minipage}
\caption{PSF of the lens made with Ge(111) without any misalignment errors
and also with no radial distortion. Each pixel is having a dimension of 200$\mu \times$200$\mu$. }
\label{fig:L_Ge_M00R0}
\end{figure}

\begin{figure}[!ht]
\centering
\begin{minipage}{.5\textwidth}
    \centering
    \includegraphics[scale=0.45]{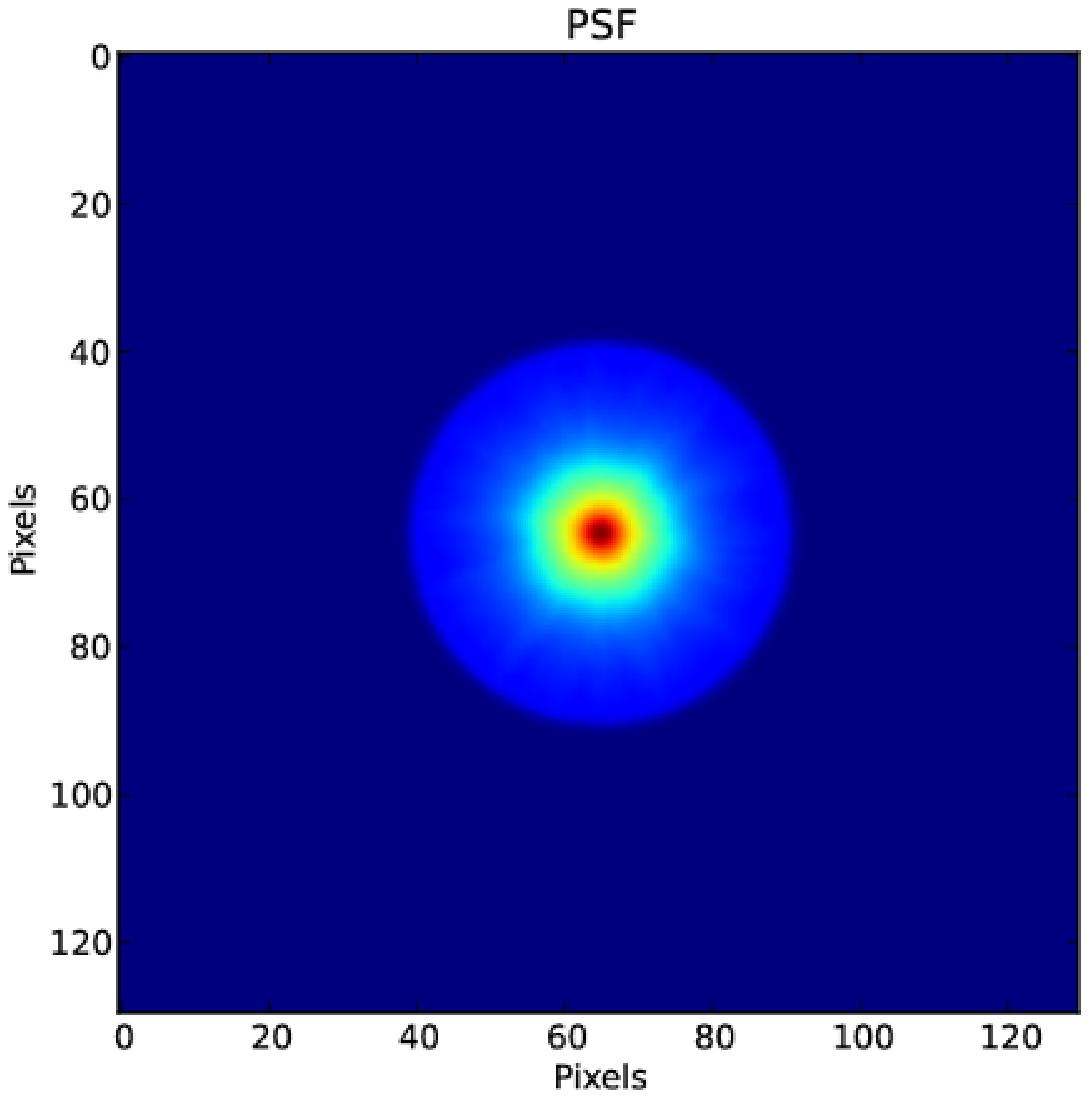}
    \renewcommand{\figure}{Fig.}
    
\end{minipage}%
\begin{minipage}{.5\textwidth}
    \centering
    \includegraphics[scale=0.65]{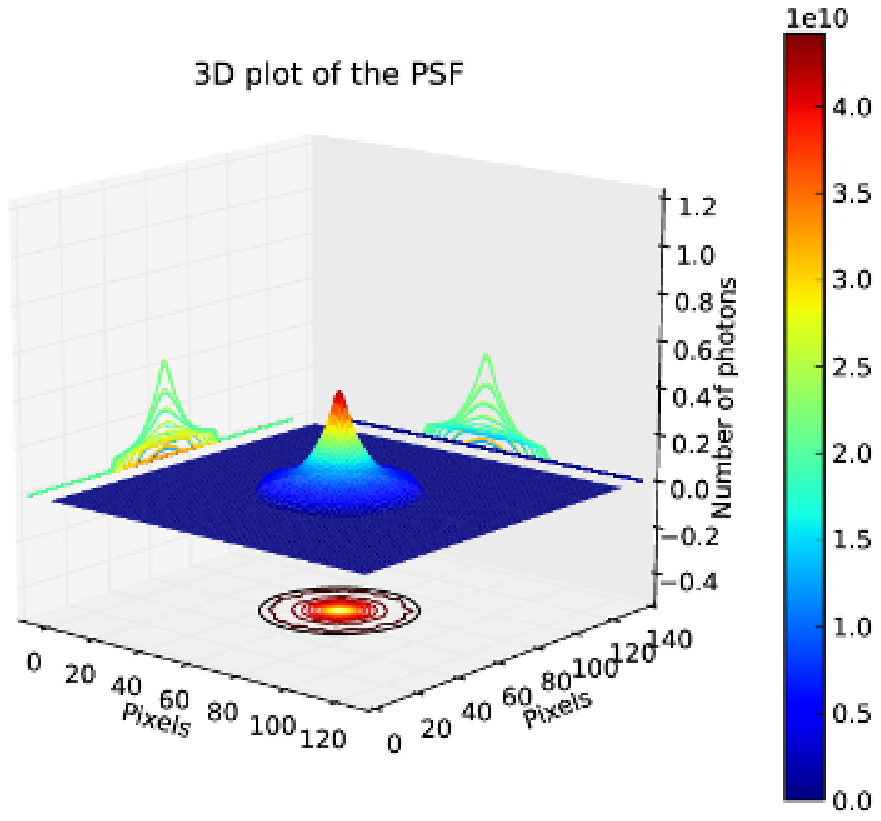} 
    \renewcommand{\figure}{Fig.}
    
\end{minipage}
\caption{PSF of the lens made with Ge(111) without any misalignment errors
but with a maximum radial distortion of 6 meters. Each pixel is having a dimension of 200$\mu \times$200$\mu$.}
\label{fig:L_Ge_M00R6}
\end{figure}

\begin{figure}[!ht]
\centering
\begin{minipage}{.5\textwidth}
    \centering
    \includegraphics[scale=0.45]{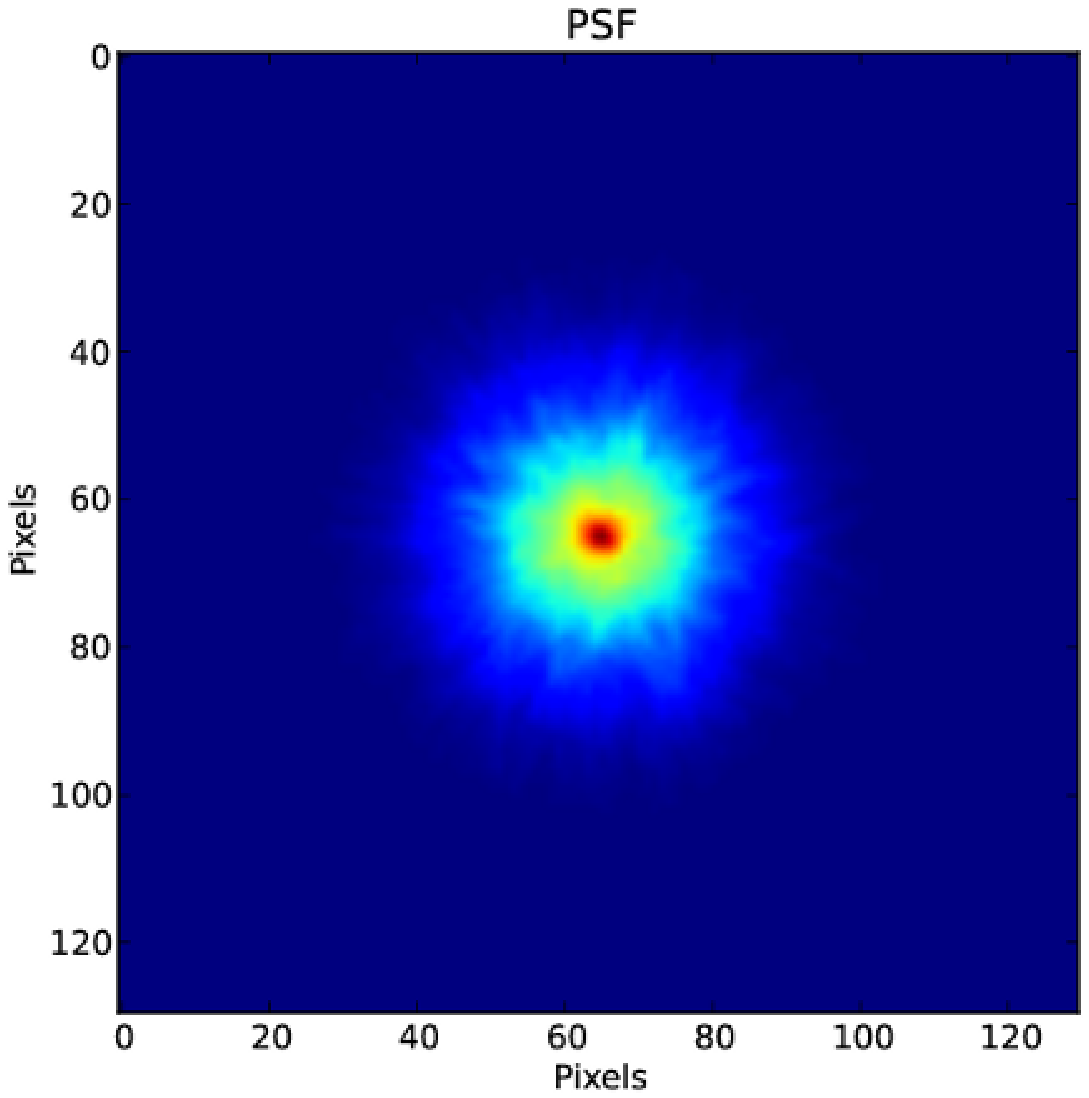}
    \renewcommand{\figure}{Fig.}
    
\end{minipage}%
\begin{minipage}{.5\textwidth}
    \centering
    \includegraphics[scale=0.65]{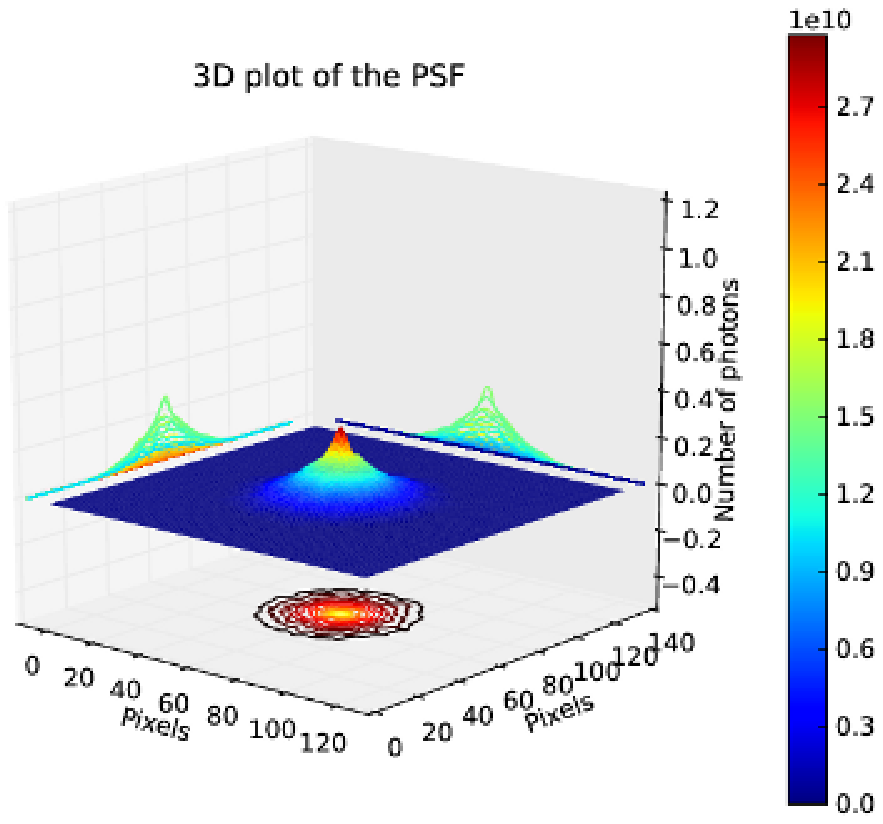} 
    \renewcommand{\figure}{Fig.}
    
\end{minipage}
\caption{PSF of the lens made with Ge(111) with a maximum misalignment of 30 arcsec
and without any radial distortion. Each pixel is having a dimension of 200$\mu \times$200$\mu$.}
\label{fig:L_Ge_M30R0}
\end{figure}

\begin{figure}[!ht]
\centering
\begin{minipage}{.5\textwidth}
    \centering
    \includegraphics[scale=0.45]{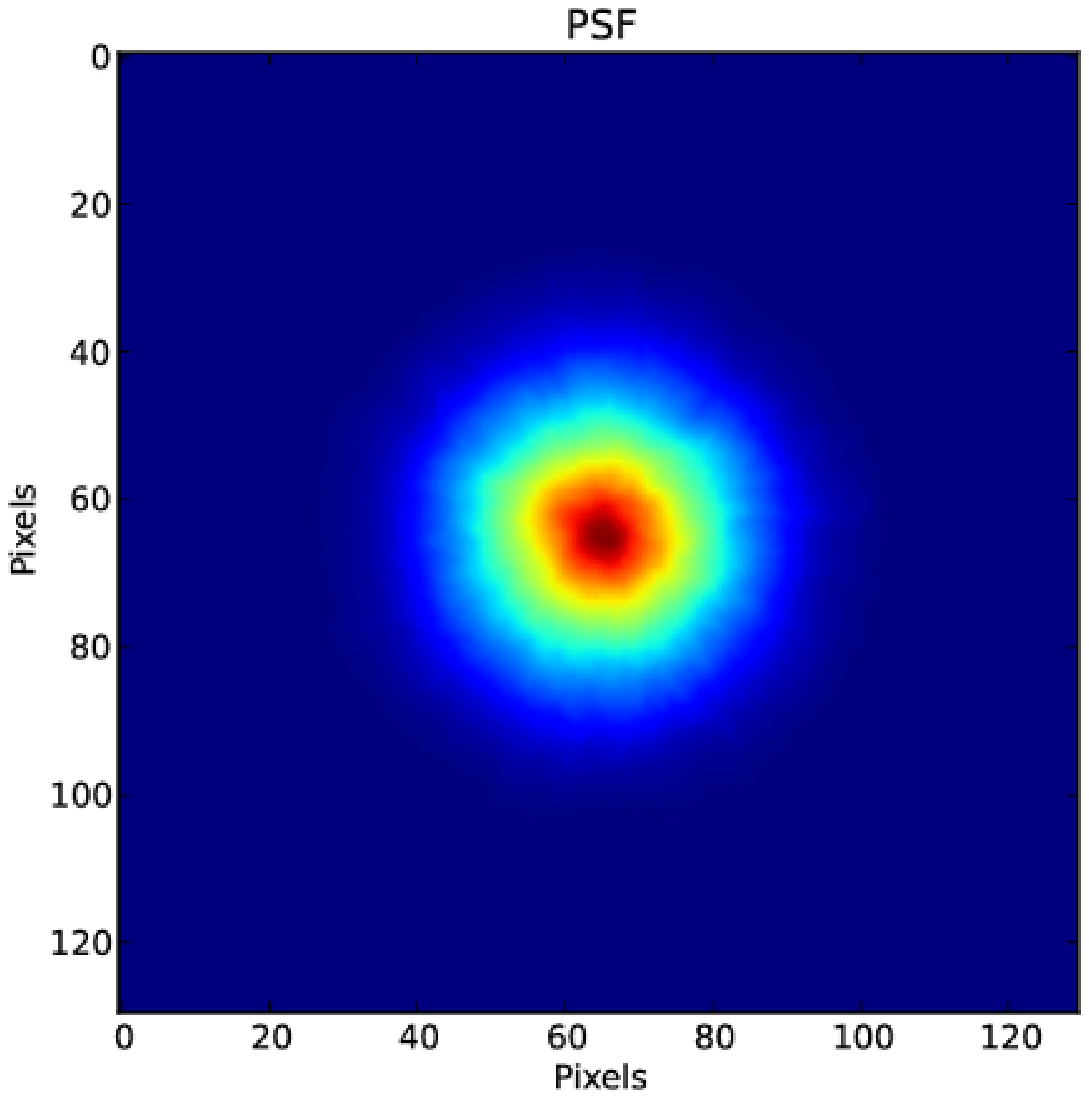}
    \renewcommand{\figure}{Fig.}
    
\end{minipage}%
\begin{minipage}{.5\textwidth}
    \centering
    \includegraphics[scale=0.65]{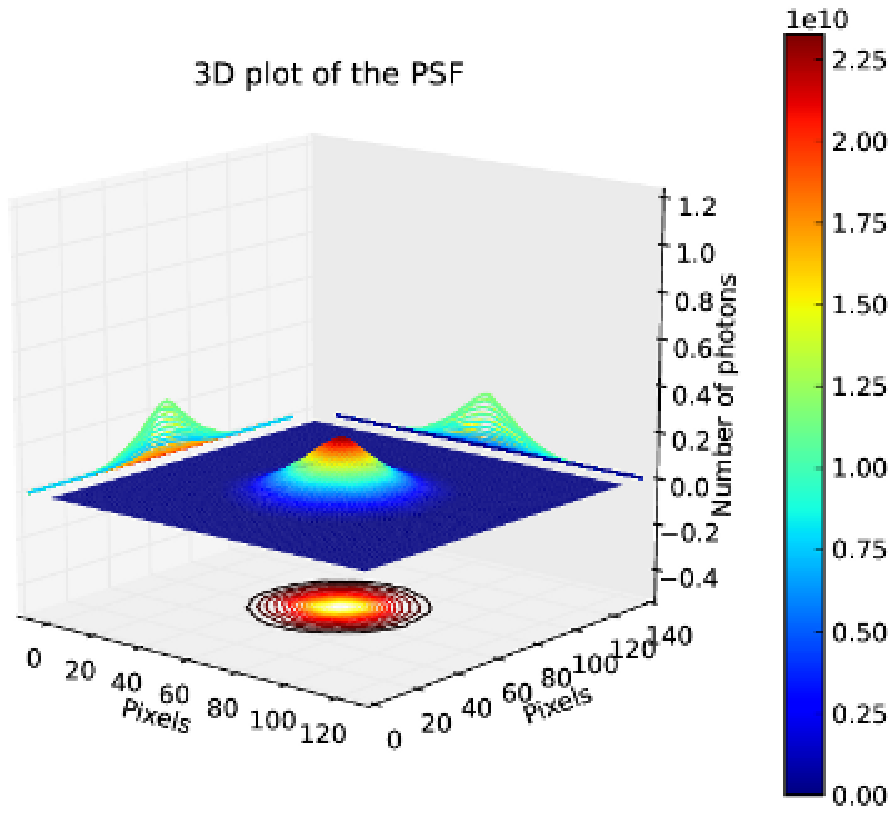} 
    \renewcommand{\figure}{Fig.}
    
\end{minipage}
\caption{PSF of the lens made with Ge(111) with a maximum misalignment of 30 arcsec
and with a maximum radial distortion of 6 meters. Each pixel is having a dimension of 200$\mu \times$200$\mu$.}
\label{fig:L_Ge_M30R6}
\end{figure}

The simulation results are shown in Fig. \ref{fig:L_Ge_M00R0}, \ref{fig:L_Ge_M00R6}, \ref{fig:L_Ge_M30R0} and 
\ref{fig:L_Ge_M30R6}. 
The energy band is from 80 keV to 600 keV.
The \textit{left} panels show the image of the PSF, while the \textit{right} panels show the 3D plot.


\newpage
\section{Quantitative analysis of the simulation results}

This section deals with the quantitative analysis of the results obtained from the simulations and their consequences.
In particular we discuss the radial profile 
of the enclosed photons, the achievable effective area, the sensitivity for both continuum and lines of the complete lens.

\subsection{Full Width at Half Maximum (FWHM) of the PSF}

\begin{figure}[h!]
  \centering
  \includegraphics[scale=0.5,keepaspectratio=true]{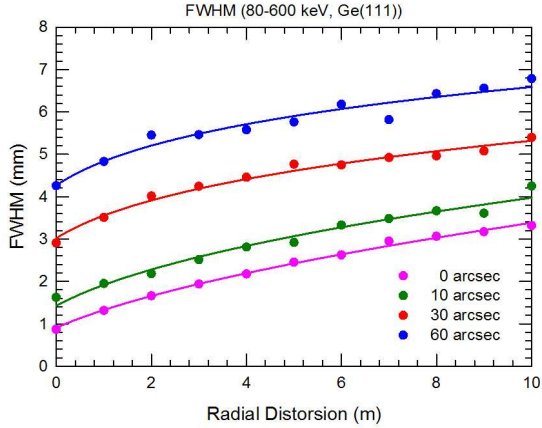}
  \renewcommand{\figure}{Fig.}
  \caption{FWHM profile of the lens made with Ge(111) for different values of crystal misalignment
  and radial distortion. The values in the legend shows maximum misalignment (in arcsec)
  in the positioning of crystals.}
  \label{fig:L_Ge_FWHM}
\end{figure}

Any misalignment in the positioning of the crystal on the lens frame
and also any distortion in the curvature radius affects the FWHM
of the PSF.
Figure~\ref{fig:L_Ge_FWHM} shows the FWHM profiles of the simulated lens as function of the radial distortion.
Each value (in meters) reported on the $x-axis$ 
corresponds to the maximum value of the uniformly distributed radial distortion, $[0,x)$ meters, 
from the ideal radius of 40 meters.  
Each line in the plot represents the corresponding value of maximum misalignment
in the positioning of the crystal tile
(shown in the $legend$ of the plot).

The model used for fitting is a power law with a threshold and having two free parameters.

\begin{equation}
 FWHM = a(1 + RD)^c
 \label{Eqn.powerLawModel}
\end{equation}

For this model the best-fit values of the parameter seems more correlated with the tiles
misalignments: the parameter $a$ reproduce directly the minimum achievable FWHM for
perfect tile alignment, while the power coefficient, $c$ decrease monotonically as the tile
misalignment level increase.

\begin{figure}[h!]
 \centering
 \includegraphics[bb=14 14 453 351,scale=0.5,keepaspectratio=true]{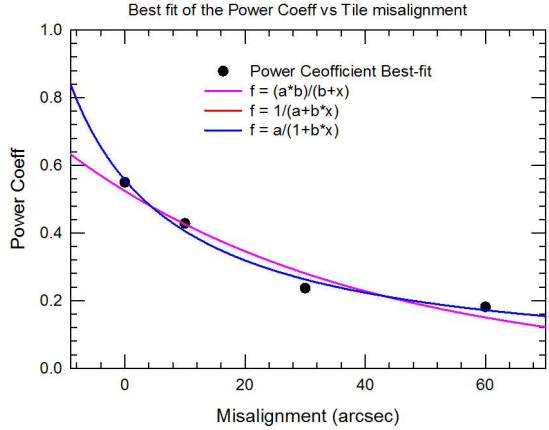}
 \caption{Best fit of the power coefficient vs tile misalignement.}
 \label{fig:power_coeff}
\end{figure}

Fig. \ref{fig:power_coeff}, shows the tentative models that has been considered 
to fit the power law coefficient of the model shown in Eqn. \ref{Eqn.powerLawModel}, 
to obtain a semi-empirical relationship to evaluate a-priori the FWHM starting from the
values expected for the case of perfect tile alignment on the lens support.
In the graph the last two function used (red line and blue line) are perfectly superimposing
and therefore, at least in the considered interval of misalignments, are equivalent. In any
case, these two model describe better the behaviour of the power coefficient as a function of
tile misalignment (correlation coefficient is $>$ 0.99 and the standard error of estimate is about
0.026).
The parameters for these last two fit are:

\begin{equation*}
 \begin{aligned}
  a = 1.80 \pm 0.08, ~& b = 0.067 \pm 0.009; ~~\textnormal{for} ~ f = 1/(a+bx) ~~\textnormal{and}\\
  a = 0.56 \pm 0.02, ~& b = 0.037 \pm 0.006; ~~\textnormal{for} ~ f = a/(1+bx)
 \end{aligned}
\end{equation*}

As can be seen from Fig.~\ref{fig:L_Ge_FWHM},
the FWHM of the PSF increases with the misalignment value 
as well as with the radial distortion. When there is no misalignment and no radial distortion,
that is, when the lens is perfectly built, the FWHM is 0.6 mm, which is that expected for a single crystal tile. 

\subsection{Peak intensity and radial profile}

\begin{figure}[h!]
  \centering
  \includegraphics[scale=0.4,keepaspectratio=true]{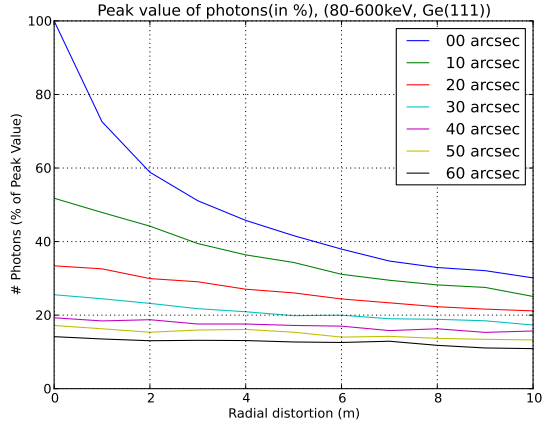}
  \renewcommand{\figure}{Fig.}
  \caption{Peak intensity profile of the petal made with Ge(111) for different values of crystal misalignments
  and radial distortion. The values in the legend shows maximum misalignment (in arcsec)
  in the positioning of crystals.}
  \label{fig:L_Ge_PI}
\end{figure}

Any misalignment in the positioning of the crystal on the lens frame
and also any distortion in the curvature radius from the required value of 40 meters 
will also affect the peak intensity of the PSF.
This effect is plotted in Fig.~\ref{fig:L_Ge_PI}.
The peak intensity, as expected, gets reduced with the increase of the crystal misalignment and
radial distortion. 
With respect to a perfect lens made of Ge(111) without any 
misalignment or radial distortion, the peak intensity gets reduced to 20\%
for a maximum misalignment of 
30 arcsec and a maximum radial distortion of 6 meters.
This is also evident from the comparison of Fig.~\ref{fig:L_Ge_M00R0}
with Fig.~\ref{fig:L_Ge_M30R6}.

\begin{figure}[h!]
 \centering
 \includegraphics[scale=0.3]{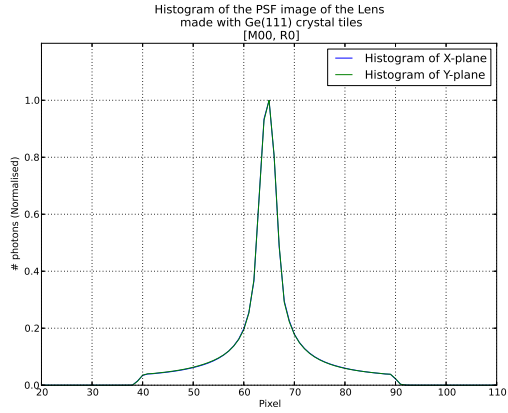} 
 \caption{PSF profile of the lens made of crystal tiles of Ge(111). 
  A perfect positioning of the crystal tiles and no radial distortions are assumed.}
  \label{fig:histo_Lens}
\end{figure}

The PSF radial profile is obtained by calculating the total photons radially enclosed outwards from the focal 
point of the PSF. The PSF profile of the lens made of Ge(111) with perfect alignment 
of the crystal tiles and no radial distortion is shown in Fig.~\ref{fig:histo_Lens}. 
The values of the radius (in both pixels and mm) for a given enclosed percentage of photons, are reported 
in the Table~\ref{tab:rad_profile_Lens}.

\begin{table}[h!]
  \begin{center}  
  \caption{Radial distribution of the enclosed photons for a 
  perfect positioning of the crystal tiles without any radial distortion.}
  \vspace{0.3 cm}
  \begin{tabular}{ c c c }
    \toprule 
    \multirow{2}{*}{Enclosed photons} & \multicolumn{2}{c}{Radius} \\ \cmidrule(r){2-3}
	(\%)	& (pixels)	& (mm)	\\  \midrule 
    
    10		& 3		& 0.6	\\  
    20		& 4		& 0.8	\\  
    30		& 5		& 1.0	\\  
    40		& 6		& 1.2	\\  
    50		& 7		& 1.4	\\  
    60		& 8		& 1.6	\\  
    70		& 10		& 2.0	\\  
    80		& 12		& 2.4	\\  
    90		& 14		& 2.8	\\  
    100		& 20		& 4.0	\\  
  \bottomrule
  \end{tabular}
  \label{tab:rad_profile_Lens}
  \end{center}	
\end{table}

\subsection{Effective area}
\label{sec:EA}

The effective area of a lens made of crystals at an energy $E$ is defined as
the product of the geometric area of the 
crystals that reflect photons in a narrow energy interval $\Delta E$ around $E$ $ \left( E - \dfrac{\Delta E}{2} \leq
 E \leq E + \dfrac{\Delta E}{2} \right) $ times the mean reflection efficiency in this energy interval.

In our case, if the entire energy range (90-600 keV) of the lens is sub--divided into 10 equal bins 
in logarithmic scale, the total geometric area in the i$^{th}$ energy bin, $G_{total}^{bin}$, is the total
cross section of the crystals that reflect photons in the energy range of the bin.

For example, if $E_{min}^{bin}$ and $E_{max}^{bin}$ are respectively the minimum and maximum energy 
of a given bin, then the total geometric area $G_{total}^{bin}$, in this energy bin is given by:

\begin{equation}
  G_{total}^{bin} = \pi (R_{max}^2 - R_{min}^2)
   \label{eqn:TGA_pi}
\end{equation}

where, through the Bragg law,
\begin{align*}
  R_{max} &= \frac{hc f}{d_{hkl} E_{min}^{bin}}\\
  R_{min} &= \frac{hc f}{d_{hkl} E_{max}^{bin}} 
\end{align*}

If $N_c(\Delta E)$ is the number of crystal tiles that reflect photons in a given energy bin,
and $A_{xtal}$ is the surface area ($dim[0] \times dim[1]$) of a single crystal tile, 
then Eqn.\ref{eqn:TGA_pi} can also be written as

\begin{equation}
  G_{total}^{bin} = N_c(\Delta E) \times A_{xtal}
\end{equation}

Hence, in this method, the total geometric area $G_{total}^{bin}$, does not depend upon the thickness as well 
as the mosaicity of the crystal tile.

The effective area $A_{eff}^{bin}$, in a given energy bin is given by:

\begin{equation}
  A_{eff}^{bin} = G_{total}^{bin} \times \Re 
  \label{eqn:EA_method2}
\end{equation}

where $\Re$ is the mean reflectivity in that bin.

The large values of effective area at lower energies and smaller values at higher energies
are due to the difference in the number of crystal tiles corresponding to those energies.
For example, the number of Ge(111) crystal tiles corresponding to lowest energy bin (in case of $\Delta E$ 
sub--divided into 10 logarithmically equal bins) are around 2092, 
but there are only 68 Ge(111) crystal tiles corresponding to the highest energy bin.

\begin{table}[h!]
  \begin{center}  
  \caption{The values of different parameters for each of the 10 logarithmic energy bins for the lens made with Ge(111) 
  crystal tiles.}
  \vspace{0.3 cm}
  \begin{tabular}{ c c c r r c r }
    \toprule 
    Energy range & $R_{max}$ & $R_{min}$ & $\Delta R$ & $G_{total}^{bin}$ & \multirow{2}{*}{$\Re$} & $A_{eff}^{bin}$ 		\\
    (keV)	 & (cm)	     & (cm)      & (cm)       & ($cm^2$)           &                      & ($cm^2$)			\\  \midrule 
      90 -- 109	 & 84.35     &	69.64	 & 14.70	& 7113.75		& 0.57394	&	4082.90		\\
      109 -- 132 & 69.64     &	57.51    & 12.13	& 4848.13	 	& 0.69485	&	3368.75		\\
      132 -- 159 & 57.51     &	47.74    & 9.76		& 3229.57		& 0.77453	&	2501.40		\\
      159 -- 193 & 47.74     &	39.33    & 8.41		& 2301.14		& 0.82582	&	1900.32		\\
      193 -- 233 & 39.33     &	32.58    & 6.75		& 1525.71		& 0.85862	&	1310.01		\\
      233 -- 281 & 32.58     &	27.01    & 5.56	     	& 1042.1		& 0.87979	&	916.83		\\
      281 -- 340 & 27.01     &	22.32    & 4.68	 	& 726.78		& 0.89458	&	650.16		\\
      340 -- 411 & 22.32     &	18.47    & 3.85		& 494.4			& 0.90532	&	447.60		\\
      411 -- 496 & 18.47     &	15.30    & 3.16		& 335.89		& 0.91055	&	305.85		\\
      496 -- 599 & 15.30     &	12.67    & 2.63		& 231.34		& 0.90075	&	208.38		\\

  \bottomrule
  \end{tabular}
  \label{tab:EA_Ge}
  \end{center}	
\end{table}

Table~\ref{tab:EA_Ge} gives the values of different parameters that have been derived for the 
effective area calculation.

\subsection{Continuum Sensitivity}

It is well known that continuum sensitivity of a focusing telescope in the observation time $T_{obs}$ and in the energy band $ \left( E - \dfrac{\Delta E}{2}, E + \dfrac{\Delta E}{2} \right) $ ,  is given by 

\begin{equation}
  I^{min}_{ft}(E) = n_{\sigma}\frac{\sqrt{B(E)} \sqrt{A_d}}{\eta_{d} f_{\epsilon} A_{eff} \sqrt{\Delta E} \sqrt{T_{obs}}} 
 \label{eqn:CS_ft}
\end{equation}

where the instrument background is assumed to be Poissonian, and

\begin{itemize}
  \item $n_\sigma$ is the number of standard deviations $\sigma$,
  \item $B(E)$ is the intensity of the instrument background (in $counts/s/cm^2/keV$) at the energy $E$;
  \item $\eta_{d}$ is the efficiency of the position sensitive detector;
  \item $f_{\epsilon}$ is the fraction of photons that is focused on the detector area $A_d ( = \pi R_{spot}^2)$;
  \item $T_{obs}$ is the observation time;
  \item $A_{eff}$ is the mean effective area in the energy band ;
  $ \left( E - \dfrac{\Delta E}{2} \leq E \leq E + \dfrac{\Delta E}{2} \right) $ 
  of the telescope optics.
\end{itemize}

\subsection{ Expected detection efficency and instrument background}

\begin{figure}[h]
 \centering
 \includegraphics[scale=0.6,keepaspectratio=true]{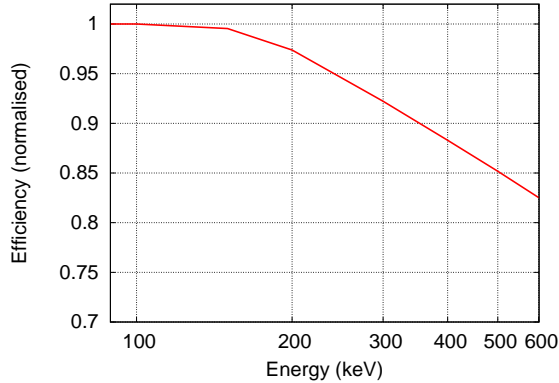}
 \renewcommand{\figure}{Fig.}
 \caption{Detection efficiency behaviour with energy of the Ge stacked detector adopted for the simulations.}
 \label{fig:det_eff}
\end{figure}

As discussed in our paper by Khalil et al. \cite{Khalil15}, 
a Germanium detector is a good candidate  
as focal plane position sensitive detector of the focused photons, with a pixel size of 350 $\mu$m $\times$ 350 $\mu$m, 
a cross section of $10\times 10$~cm$^2$ and a thickness of 12.5 cm or higher. 
In such a way a high detection efficiency (80\%) can be achieved up to the highest energies of the lens pass band (650 keV).
The detector efficiency $\eta_d$ behaviour adopted for the simulations is shown in Fig.~\ref{fig:det_eff}. 

\begin{figure}[h]
 \centering
 \includegraphics[bb=50 50 410 302,scale=0.7,keepaspectratio=true]{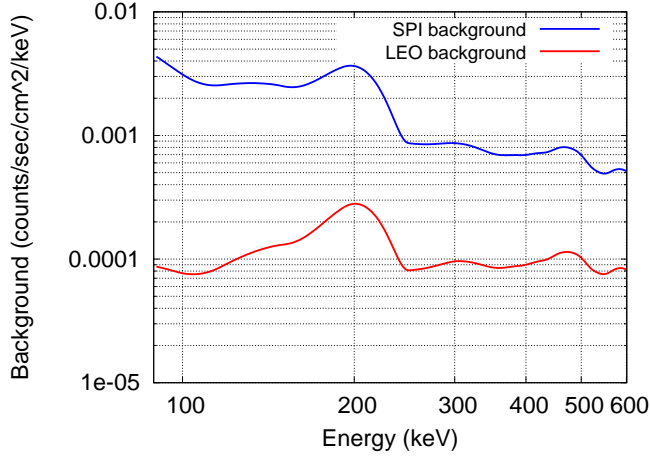}
 \caption{The expected lens background at LEO, compared with the INTEGRAL SPI measured background.}
 \label{fig:bkgd}
\end{figure}

Concerning the instrument backgound, for our simulations the Laue lens is assumed to be placed in a Low Earth Orbit (LEO).
Following the same procedure discussed in our paper by Khalil et al.\cite{Khalil15}, from the background data in the energy band from 90 -- 600 keV 
measured by SPI instrument onboard the INTEGRAL satellite in High Earth Orbit, we have derived the background level expected for the lens. 
The result is shown in Fig.~\ref{fig:bkgd}. 
In addition to the expected lens background, the SPI measured background is shown, for comparison.

\begin{figure}[h!]
 \centering
 \includegraphics[angle=270,scale=0.5,keepaspectratio=true]{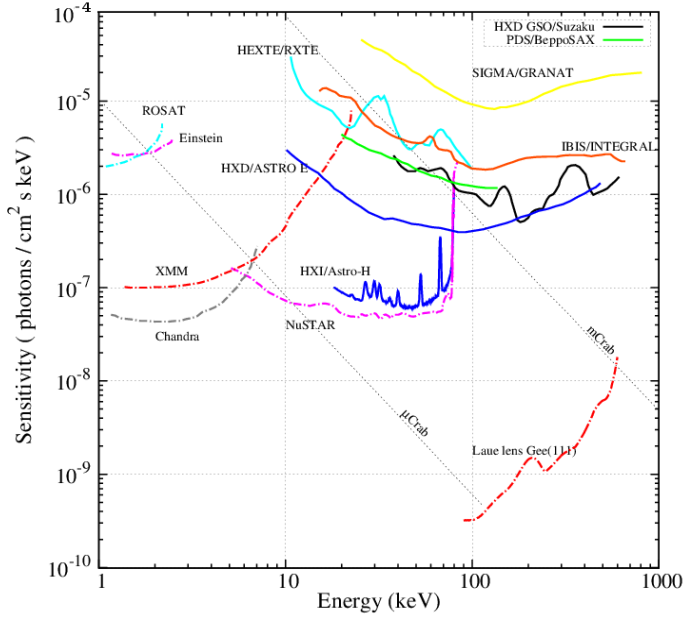}
 \caption{Sensitivity of different missions. Continuous line shows the direct-view instruments\//missions. 
 Dashed line depicts the focusing instruments\//missions.}
 \label{fig:SensitivityMissions}
\end{figure}

By inserting in the Eqn.~\ref{eqn:CS_ft} the background level and  detection efficiency given above, we can get the
continuum sensitivity of the simulated Laue lens.
Figure \ref{fig:SensitivityMissions} shows the result, at 3$\sigma$ level for an observation time $T_{obs} = 10^5$s and $\Delta E = E/2$, 
along with the continuum sensitivity  of other missions.

From this figure it is apparent that at 100 keV,
the simulated lens is 4 orders of magnitude more sensitive than INTEGRAL IBIS. Similar results are obtained when we compare the lens 
with the other INTEGRAL instrument SPI.
The data of ISGRI and SPI are taken from the reference \cite{Belanger12} and \cite{Fernandez12}, respectively.
At about 511 keV, for a lens made with Ge(111), the continuous sensitivity is more than 2 orders of magnitude better 
than IBIS.

\subsection{Sensitivity to narrow emission lines}

The sensitivity to narrow emission lines is derived by superposing the emission line to the continuum source level.
In the case of a lens telescope, if the source continuum level can be accurately determined,
the minimum detectable intensity $I_L^{min}$ (in $photons/s/cm^2$),
of a line is given by 

\begin{equation}
  I_L^{min}(E_L) = 1.31 n_{\sigma}\frac{\sqrt{[2 B(E_L) A_d + I_c(E_L) \eta_{d} f_{\epsilon} A_{eff}]\Delta E} } { \eta_{d} f_{\epsilon} A_{eff} \sqrt{T_{obs}}} 
 \label{eqn:LS_ft}
\end{equation}

where $B(E_L)$ is the intensity of the measured background spectrum (in counts/s/cm$^2$/keV) at the line energy $E_L$; 
$I_c(E_L)$ is the source continuum intensity (in photons/s/cm$^2$/keV) at the centroid of the line, 
$A_{eff}$ is the lens effective area at an energy $E_L$, $A_d$ is the detection area that encloses $f_{\epsilon} = $ 50\% of the 
focused photons (half power), and
$\Delta E$ is the FWHM of the line profile around $E_L$. 
The  $\Delta E$ value depends upon the energy resolution of the detector. 
We have assumed $\Delta E$ = 2 keV (expected for our simulated detector), detector area $A_d$ corresponding to the half power radius 
given in Table \ref{tab:rad_profile_Lens}, $\eta_{d}$ shown in Fig.~\ref{fig:det_eff},
and  $A_{eff}$ derived from Eqn. \ref{eqn:EA_method2} with a filling factor of 0.2 mm.

The result at different energies and for an observation time of $10^5$~s and $10^6$~s  is shown in Table~\ref{tab:LS_Lens}.

\begin{table}[!h]
  \begin{center}  
  \caption{$3 \sigma$ line sensitivity (in photons/s/cm$^2$) of a lens with time of observation times of 
  $10^5$ and $10^6$.}
  \vspace{0.3 cm}
  \begin{tabular}{ c c c c c }
    \toprule 
    Energy & $T_{obs} = 10^5 sec$	& $T_{obs} = 10^6 sec$ \\ 
    (keV)	& 		& 				\\  \midrule 
    100		& $1.08 \times 10^{-8}$		&  $3.26 \times 10^{-9}$				\\  
    200		&  $3.31 \times 10^{-8}$	&  $1.04 \times 10^{-8}$				\\  
    300		&  $7.19 \times 10^{-8}$	&  $2.27 \times 10^{-8}$				\\  
    400		&  $1.47 \times 10^{-7}$	&  $4.67 \times 10^{-8}$				\\  
    511		&  $2.39 \times 10^{-7}$	&  $7.55 \times 10^{-8}$				\\  
  \bottomrule
  \end{tabular}
  \label{tab:LS_Lens}
  \end{center}	
\end{table}

\begin{table}[!h]
  \begin{center}  
  \caption{$3 \sigma$ line sensitivity (in photons/s/cm$^2$) of ISGRI 
  and of SPI (observation time = $10^6$ seconds, $\Delta E = E/2$). }
  \vspace{0.3 cm}
  \begin{tabular}{ c c c c}
    \toprule 
    \multicolumn{2}{c}{ISGRI}			& \multicolumn{2}{c}{SPI}	\\ \cmidrule(r){1-2} \cmidrule(r){3-4}
    Energy	&  Line sensitivity		& Energy	& Line sensitivity 	\\
    (keV)	& (photons/s/cm$^2$)		& (keV)		& (photons/s/cm$^2$) 	\\ \midrule
    112.9	& $1.67 \times 10^{-5}$		& 100		& $4.4 \times 10^{-5}$		\\
    225.4	& $3.21 \times 10^{-5}$	 	& --		& --				\\
    449.6	& $6.64 \times 10^{-5}$		& --		& --				\\
    566.1	& $8.40 \times 10^{-5}$		& 500		& $3.1 \times 10^{-5}$		\\
  \bottomrule
  \end{tabular}
  \label{tab:LS_integral}
  \end{center}	
\end{table}

For comparison, in Table \ref{tab:LS_integral} we report the line sensitivity at different energies and for an observation time of $10^6$~s 
shown by IBIS and SPI aboard INTEGRAL. It can be seen that the lens expected sensitivity, at about 100 keV, is 4 orders of magnitude better, 
and,  at 511 keV, is 3 orders of magnitude better.

\section{Conclusions} 

A complete Laue lens made of Ge(111) crystal tiles is simulated and modeled, with an 
energy passband from 90 -- 600 keV. The on-axis PSF, its FWHM,
effective area and sensitivity (to continuum and to narrow emission lines) of the entire Laue 
lens have been modeled.  The sensitivity results show that this lens is almost 3 orders of 
magnitude more sensitive than ISGRI \cite{Belanger12} and SPI \cite{Fernandez12} on-board the INTEGRAL satellite.

With the sensitivity that this lens is expected it can achieve, many fundamental open 
astrophysical cases \cite{Frontera13} can be settled.

\begin{acknowledgements}
Vineeth Valsan acknowledges the support from Erasmus Mundus Joint Doctorate Program by Grant Number 2010-1816 from 
the EACEA of the European Commission. The LAUE project is the result of huge efforts of large number of organizations and people. We would like
to thank all of them. We also acknowledge the ASI (Italian Space Agency) for its support to the LAUE project
under contract I/068/09/0. Currently, V. Valsan is supported by India-TMT Project.
\end{acknowledgements}

\bibliographystyle{spphys} 
\bibliography{bib_ref}

\end{document}